\RequirePackage{ifpdf}
\ifpdf 
\documentclass[pdftex]{sigma}
\else
\documentclass{sigma}
\fi

\numberwithin{equation}{section}

\begin{document}

\allowdisplaybreaks

\renewcommand{\thefootnote}{$\star$}

\renewcommand{\PaperNumber}{035}

\FirstPageHeading

\newcommand{\kb}{\mathbf{k}}
\newcommand{\thetab}{\boldsymbol{\theta}}
\newcommand{\lambdab}{\boldsymbol{\lambda}}
\newcommand{\pb}{\mathbf{p}}
\newcommand{\nb}{\mathbf{n}}
\newcommand{\phib}{\boldsymbol{\phi}}
\newcommand{\field}[1]{\mathbb{#1}}
\newcommand{\R}{\field{R}}
\newcommand{\Z}{\field{Z}}
\newcommand{\N}{\field{N}}
\newcommand{\be}{\bar{E}}
\newcommand{\bpsi}{\bar{\Psi}}
\newcommand{\ce}{{\cal E}}
\newcommand{\ca}{{\cal A}}
\newcommand{\ct}{{\cal T}}
\newcommand{\cd}{{\cal D}}

\ShortArticleName{Revisiting the Symmetries of the Quantum Smorodinsky--Winternitz System}

\ArticleName{Revisiting the Symmetries of the Quantum\\ Smorodinsky--Winternitz System in $\boldsymbol{D}$ Dimensions\footnote{This paper is a
contribution to the Special Issue ``Symmetry, Separation, Super-integrability and Special Functions~(S$^4$)''. The
full collection is available at
\href{http://www.emis.de/journals/SIGMA/S4.html}{http://www.emis.de/journals/SIGMA/S4.html}}}

\Author{Christiane QUESNE}

\AuthorNameForHeading{C.~Quesne}

\Address{Physique Nucl\'eaire Th\'eorique et Physique Math\'ematique,  Universit\'e Libre de
Bruxelles, \\ Campus de la Plaine CP229, Boulevard~du Triomphe, B-1050 Brussels,
Belgium}
\Email{\href{mailto:cquesne@ulb.ac.be}{cquesne@ulb.ac.be}}

\ArticleDates{Received January 17, 2011, in f\/inal form March 25, 2011;  Published online April 02, 2011}

\Abstract{The $D$-dimensional Smorodinsky--Winternitz system, proposed some years ago by Evans, is re-examined from an algebraic viewpoint. It is shown to possess a potential algebra, as well as a dynamical potential one, in addition to its known symmetry and dynamical algebras. The f\/irst two are obtained in hyperspherical coordinates by introdu\-cing~$D$ auxiliary continuous variables and by reducing a $2D$-dimensional harmonic oscillator Hamiltonian. The su($2D$) symmetry and ${\rm w}(2D) \oplus_s {\rm sp}(4D, \R)$ dynamical algebras of this Hamiltonian are then transformed into the searched for potential and dynamical potential algebras of the Smorodinsky--Winternitz system. The action of generators on wavefunctions is given in explicit form for $D=2$.}


\Classification{20C35; 81R05; 81R12}

\renewcommand{\thefootnote}{\arabic{footnote}}
\setcounter{footnote}{0}

\section{Introduction}

In classical mechanics, a Hamiltonian $H$ with $D$ degrees of freedom is said to be completely integrable if it allows $D$ integrals of motion $X_{\mu}$, $\mu = 1, 2, \ldots, D$, that are well-def\/ined functions on phase space, are in involution and are functionally independent (see, e.g.,~\cite{goldstein}). These include the Hamiltonian, so that we may assume $X_D = H$. The system is superintegrable if there exist~$k$ additional integrals of motion $Y_{\nu}$, $\nu=1, 2, \ldots, k$, $1 \le k \le D-1$, that are also well-def\/ined functions on phase space and are such that the integrals $H$, $X_1, X_2, \ldots, X_{D-1}$, $Y_1, Y_2, \ldots, Y_k$ are functionally independent. The cases $k=1$ and $k=D-1$ correspond to minimal and maximal superintegrability, respectively.

Similar def\/initions apply in quantum mechanics with Poisson brackets replaced by commutators, but $H$, $X_{\mu}$, and $Y_{\nu}$ must now be well-def\/ined operators forming an algebraically independent set. Maximally superintegrable quantum systems appear in many domains of physics, such as condensed matter as well as atomic, molecular, and nuclear physics. They have a lot of nice properties: they can be exactly (or quasi-exactly) solved, they are often separable in several coordinate systems and their spectrum presents some ``accidental'' degeneracies, i.e., degeneracies that do not follow from the geometrical symmetries of the problem.

The most familiar examples of such systems are the Kepler--Coulomb \cite{pauli, fock, bargmann} and the oscilla\-tor~\cite{jauch, moshinsky} ones. Other well-known instances are those resulting from the f\/irst systematic search for superintegrable Hamiltonians on~E$_2$ carried out by Smorodinsky, Winternitz, and collabora\-tors~\cite{fris, winternitz, makharov} and from its continuation by Evans on~E$_3$~\cite{evans90a}. These studies were restricted to those cases where the integrals of motion are f\/irst- or second-order polynomials in the momenta. Later on, many ef\/forts have been devoted to arriving at a complete classif\/ication of these so-called second-order superintegrable systems (see, e.g., \cite{kalnins05a, kalnins05b, kalnins05c, kalnins06a, kalnins06b, daska06, daska07}).

Only recently, the pioneering work of Drach \cite{drach35a, drach35b} on two-dimensional Hamiltonian systems with third-order integrals of motion has been continued \cite{gravel02, gravel04}. Nowadays the search for $D$-dimensional superintegrable systems with higher-order integrals of motion has become a very active f\/ield of research (see, e.g., \cite{verrier, evans08, rodriguez, marquette09a, marquette09b, tremblay, cq10, kalnins10a, kalnins10b, post, kalnins10c}).

In the present paper, we plan to re-examine from an algebraic viewpoint one of the classical examples of $D$-dimensional superintegrable quantum systems, namely the Smorodinsky--Winternitz (SW) one \cite{fris, winternitz, makharov, evans90a, evans90b, evans91}, which may be def\/ined in Cartesian coordinates as
\begin{gather}
  H^{(\kb)} = \sum_{\mu=1}^D \left(- \partial_{x_{\mu}}^2 + \frac{k_{\mu}^2}{x_{\mu}^2} + \omega^2
  x_{\mu}^2\right).  \label{eq:H-SW}
\end{gather}
Here $\omega$, $k_1, k_2, \ldots, k_D$ are some constants, which we assume to be real and positive.

Several distinct algebraic methods may be used in connection with superintegrable systems. One of them is based on the fact that the integrals of motion generate a nonlinear algebra closing at some order \cite{granovskii92a, granovskii92b, granovskii92c}. It has been shown, for instance, that for two-dimensional second-order superintegrable systems with nondegenerate potential and the corresponding three-dimensional conformally f\/lat systems, one gets a quadratic algebra closing at order 6 \cite{kalnins05a, kalnins05b, kalnins05c, kalnins06a, kalnins06b}. Its f\/inite-dimensional unitary representations can be determined \cite{daska01} by using a deformed parafermion oscillator realization \cite{daska91, cq94}, thereby allowing a calculation of the energy spectrum. This procedure can be extended to higher-order integrals of motion and to the corresponding higher-degree nonlinear algebras \cite{marquette09a, marquette09b}.

Superintegrable systems may also be related \cite{marquette09a, marquette09b} to systems studied in supersymmetric quantum mechanics \cite{cooper, junker} or higher-order supersymmetric quantum mechanics \cite{andrianov95a, andrianov95b, andrianov95c, samsonov, bagchi, plyu, klishevich, aoyama, fernandez}, hence can be described in terms of either linear or nonlinear superalgebras. As a consequence, supersymmetry provides a convenient tool for generating superintegrable quantum systems with higher-order integrals of motion \cite{marquette09c, marquette10}.

The concept of exact or quasi-exact solvability \cite{turbiner, shifman, ushve}, based on the existence of an inf\/inite f\/lag of functionally linear spaces preserved by the Hamiltonian or only that of one of these spaces, appears to be related to f\/inite-dimensional representations of some Lie algebras of f\/irst-order dif\/ferential operators, such as sl(2,$\R$), sl(3,$\R$), etc. Although dif\/ferent from the concept of superintegrability, it can be related to the latter for some superintegrable systems (see, e.g., \cite{tremblay, klishevich, tempesta}). It is worth noting, however, that some alternative def\/initions of exact and quasi-exact solvability have been proposed for some specif\/ic superintegrable systems in connection with multiseparability of the corresponding Schr\"odinger equation \cite{kalnins06c, kalnins07}.

The accidental degeneracies appearing in the bound-state spectrum of superintegrable quantum systems may be understood in terms of a symmetry algebra, which is such that for any energy level the wavefunctions corresponding to degenerate states span the carrier space of one of its unitary irreducible representations (unirreps) \cite{demkov, dothan70}. The generators of this symmetry algebra, commuting with the Hamiltonian, are integrals of motion, which may assume a rather complicated form in terms of some basic ones due to the fact that linear algebras are often preferred\footnote{It is worth observing here that this may be seen as the obverse of the approach used in \cite{kalnins05a, kalnins05b, kalnins05c, kalnins06a, kalnins06b, granovskii92a, granovskii92b, granovskii92c, daska01}, where the generators are the basic integrals of motion but the algebra turns out to be nonlinear.} (note, however, that nonlinear algebras may also be considered \cite{bonatsos}). A familiar examp\-le of this phenomenon is provided by the so(4) symmetry algebra of the three-dimensional Kepler--Coulomb problem \cite{pauli, fock, bargmann}. Another one corresponds to the su(3) symmetry algebra of the three-dimensional SW system \cite{evans91} (or, in general, su($D$) for the $D$-dimensional one).

In some cases, the symmetry algebra can be enlarged to a spectrum generating algebra (also called dynamical algebra) by including some ladder operators, which are not integrals of motion but act as raising or lowering operators on the bound-state wavefunctions in such a way that all of them carry a single unirrep of the algebra \cite{dothan65, mukunda, barut}. For the three-dimensional SW system, it has been shown \cite{evans91} to be given by the semidirect sum Lie algebra ${\rm w}(3) \oplus_s {\rm sp}(6, \R)$, where w(3) denotes a Weyl algebra (or, in general, by ${\rm w}(D) \oplus_s {\rm sp}(2D, \R)$ in $D$ dimensions).

For one-dimensional systems, three other types of Lie algebraic approaches have been extensively studied. All of them rely on an embedding of the system into a higher-dimensional space by introducing some auxiliary continuous variables and on the subsequent reduction of the extended system to the initial one, a procedure also used in discussing superintegrability (see, e.g., \cite{rodriguez}). They work for hierarchies of Hamiltonians, whose members correspond to the same potential but dif\/ferent quantized strengths. The simplest ones are the potential algebras \cite{alhassid83, alhassid86, frank}, whose unirrep carrier spaces are spanned by wavefunctions with the same energy, but dif\/ferent potential strengths. Larger algebras, which also contain some generators con\-nec\-ting wavefunctions with dif\/ferent energies, are called dynamical potential algebras \cite{cq88a, cq88b, cq89}\footnote{Some authors prefer to use the terminology of dynamical algebra of the hierarchy instead of dynamical potential algebra and to employ discrete variables, related to the quantum numbers characterizing the system, instead of continuous auxiliary variables. In this way, they get discrete-dif\/ferential realizations of the algebras~\cite{kuru}. Other authors favour the use of nonlinear superalgebras~\cite{correa}.}. Finally, a third kind of algebras, termed satellite algebras \cite{delsol98, delsol00}, have the property that there is a conserved quantity dif\/ferent from the energy.

Up to now, only the f\/irst one of these Lie algebraic approaches, namely that of potential algebras, has been applied to some $D$-dimensional superintegrable systems \cite{kerimov06a, kerimov06b, kerimov07a, kerimov07b, kerimov10, calzada06, calzada08, calzada09}.

The purpose of the present paper is threefold: f\/irst to apply this technique to the $D$-dimensional SW Hamiltonian~(\ref{eq:H-SW}), second to present for the same the f\/irst construction of a dynamical potential algebra in more than one dimension, and third to show very explicitly the action of both the potential and dynamical potential algebra generators on the wavefunctions in the two-dimensional case.

The paper is organized as follows. In Section~\ref{section2}, the solutions, as well as the symmetry and dynamical algebras, of a $2D$-dimensional harmonic oscillator are obtained in a suitable orthogonal coordinate system. In Section~\ref{section3}, they are transformed into the solutions, as well as the potential and dynamical potential algebras, of the $D$-dimensional SW system in hyperspherical coordinates. The $D=2$ case is then dealt with in detail in Section~\ref{section4}. Finally, Section~\ref{section5} contains the conclusion.

\section[$2D$-dimensional harmonic oscillator]{$\boldsymbol{2D}$-dimensional harmonic oscillator}\label{section2}

Let us consider a harmonic oscillator Hamiltonian
\begin{gather*}
  H^{\rm osc} = \sum_{\mu=1}^{2D} \big(- \partial^2_{X_{\mu}} + X_{\mu}^2\big)
\end{gather*}
in a $2D$-dimensional space, whose Cartesian coordinates are denoted by $X_{\mu}$, $\mu=1, 2, \ldots,2D$. For our purposes, it is convenient to consider it in a dif\/ferent orthogonal coordinate system, which we will now proceed to introduce.

\subsection[Harmonic oscillator in variables $R$, $\theta_1, \theta_2, \ldots, \theta_{D-1}$, $\lambda_1, \lambda_2, \ldots, \lambda_D$]{Harmonic oscillator in variables $\boldsymbol{R}$, $\boldsymbol{\theta_1, \theta_2, \ldots, \theta_{D-1}}$, $\boldsymbol{\lambda_1, \lambda_2, \ldots, \lambda_D}$}

On making the change of variables
\begin{gather}
 X_1 = R \sin \theta_1 \sin \theta_2 \cdots \sin \theta_{D-1} \sin \lambda_1, \qquad X_2 = R \sin \theta_1
        \sin \theta_2 \cdots \sin \theta_{D-1} \cos \lambda_1,\nonumber \\
 X_{2\nu-1} = R \sin \theta_1 \sin \theta_2 \cdots \sin \theta_{D-\nu} \cos \theta_{D-\nu+1}
        \sin \lambda_{\nu}, \qquad \nu=2, 3, \ldots, D-1, \nonumber\\
  X_{2\nu} = R \sin \theta_1 \sin \theta_2 \cdots \sin \theta_{D-\nu} \cos \theta_{D-\nu+1}
        \cos \lambda_{\nu}, \qquad \nu=2, 3, \ldots, D-1, \nonumber\\
  X_{2D-1} = R \cos \theta_1 \sin \lambda_D, \qquad X_{2D} = R \cos \theta_1 \cos \lambda_D,
  \label{eq:variables}
\end{gather}
where $0 \le R < \infty$, $0 \le \theta_{\nu} < \frac{\pi}{2}$, $\nu=1, 2, \ldots, D-1$, and $0 \le \lambda_{\nu} < 2\pi$, $\nu=1, 2, \ldots, D$, $H^{\rm osc}$ can be rewritten as
\begin{gather*}
  H^{\rm osc}  = - \partial_R^2 - \frac{2D-1}{R} \partial_R - \frac{1}{R^2} \Biggl\{\partial_{\theta_1}^2 +
        [(2D-3) \cot \theta_1 - \tan \theta_1] \partial_{\theta_1} \\
\phantom{H^{\rm osc}  =}{} + \sum_{\nu=2}^{D-1} \frac{1}{\sin^2 \theta_1 \sin^2 \theta_2 \cdots \sin^2 \theta_{\nu-1}}
        \Bigl[\partial_{\theta_{\nu}}^2 + [(2D-2\nu-1) \cot \theta_{\nu} - \tan \theta_{\nu}]
        \partial_{\theta_{\nu}}\Bigr] \\
\phantom{H^{\rm osc}  =}{} + \frac{1}{\sin^2 \theta_1 \sin^2 \theta_2 \cdots \sin^2 \theta_{D-1}} \partial_{\lambda_1}^2 +
        \sum_{\nu=2}^{D-1} \frac{1}{\sin^2 \theta_1 \sin^2 \theta_2 \cdots \sin^2 \theta_{D-\nu}
        \cos^2 \theta_{D-\nu+1}} \partial_{\lambda_{\nu}}^2 \\
\phantom{H^{\rm osc}  =}{} + \frac{1}{\cos^2 \theta_1} \partial_{\lambda_D}^2\Biggr\} + R^2
\end{gather*}
and is clearly separable.

In the corresponding Schr\"odinger equation
\begin{gather}
  H^{\rm osc} \Psi^{\rm osc}(R, \thetab, \lambdab) = E^{\rm osc} \Psi^{\rm osc}(R, \thetab, \lambdab)
  \label{eq:SE-osc}
\end{gather}
with $\thetab = \theta_1 \theta_2 \cdots \theta_{D-1}$ and $\lambdab = \lambda_1 \lambda_2 \cdots \lambda_D$, we may therefore write
\begin{gather}
  \Psi^{\rm osc}(R, \thetab, \lambdab) = {\cal N}^{\rm osc} {\cal L}(z) \left(\prod_{\nu=1}^{D-1}
  \Theta_{\nu}(\theta_{\nu})\right) \left(\prod_{\nu=1}^D e^{{\rm i} p_{D-\nu+1} \lambda_{\nu}}\right),
  \qquad z = R^2,  \label{eq:wf-osc}
\end{gather}
where
\begin{gather*}
  \partial_{\lambda_{\nu}}^2 \Psi^{\rm osc}(R, \thetab, \lambdab) = - p_{D-\nu+1}^2 \Psi^{\rm osc}(R, \thetab,
  \lambdab)
\end{gather*}
and $p_1, p_2, \ldots, p_D \in \Z$. The normalization constant ${\cal N}^{\rm osc}$ in (\ref{eq:wf-osc}) will be determined in such a~way that
\begin{gather}
  \int dV\, \left|\Psi^{\rm osc}(R, \thetab, \lambdab)\right|^2 = 1,  \label{eq:sp-osc}
\end{gather}
where
\begin{gather}
  dV= \prod_{\mu=1}^{2D} dX_{\mu} = R^{2D-1} dR \left[\prod_{\nu=1}^{D-1} (\sin \theta_{\nu})^{2D-2\nu-1}
  \cos \theta_{\nu} d\theta_{\nu}\right] \left(\prod_{\nu=1}^D d\lambda_{\nu}\right).  \label{eq:dV}
\end{gather}

As shown in the appendix, the angular part of wavefunctions (\ref{eq:wf-osc}) can be written as
\begin{gather}
  \Theta^{(\pb)}_{\nb}(\thetab) = \prod_{\nu=1}^{D-1} \Theta^{(a_{\nu}, b_{\nu})}_{n_{\nu}}(\theta_{\nu}),
      \qquad \nb = n_1 n_2 \cdots n_{D-1}, \qquad \pb = p_1 p_2 \cdots p_D,  \label{eq:theta-bis} \\
   \Theta^{(a_{\nu}, b_{\nu})}_{n_{\nu}}(\theta_{\nu}) = (\cos \theta_{\nu})^{a_{\nu} - \frac{1}{2}}
  (\sin \theta_{\nu})^{b_{\nu} - \frac{1}{2}} P_{n_{\nu}}^{\left(a_{\nu} - \frac{1}{2}, b_{\nu} + D - \nu -
  \frac{3}{2}\right)}(- \cos 2\theta_{\nu}),  \label{eq:theta}
\end{gather}
where $n_1, n_2, \ldots, n_{D-1} \in \N$,
\begin{gather}
  a_{\nu} = |p_{\nu}| + \tfrac{1}{2}, \qquad \nu = 1, 2, \ldots, D-1,\nonumber \\
 b_{\nu} = 2n_{\nu+1}\! + 2n_{\nu+2}\! + \cdots +2n_{D-1}\! + |p_{\nu+1}| + |p_{\nu+2}| + \cdots + |p_D| +
       \tfrac{1}{2}, \quad  \nu=1, 2, \ldots, D-2, \nonumber\\
 b_{D-1} = |p_D| + \tfrac{1}{2},
  \label{eq:a-b}
\end{gather}
and $P_{n_{\nu}}^{\left(a_{\nu} - \frac{1}{2}, b_{\nu} + D - \nu - \frac{3}{2}\right)}(- \cos 2\theta_{\nu})$
denotes a Jacobi polynomial~\cite{gradshteyn}, while the radial part can be expressed as
\begin{gather}
  {\cal L}^{(j)}_{n_r}(z) = z^j L_{n_r}^{(2j+D-1)}(z) e^{- \frac{1}{2} z},  \label{eq:L}
\end{gather}
in terms of a Laguerre polynomial \cite{gradshteyn}. Here $n_r \in \N$, while $j$ is def\/ined by
\begin{gather}
  j = n_1 + n_2 + \cdots + n_{D-1} + \tfrac{1}{2}(|p_1| + |p_2| + \cdots + |p_D|)  \label{eq:j}
\end{gather}
and may take nonnegative integer or half-integer values.

The corresponding energy eigenvalues are given by
\begin{gather}
  E^{\rm osc}_{n_r j} = 2(2n_r + 2j + D).  \label{eq:E-osc}
\end{gather}
We therefore recover the well-known spectrum of the $2D$-dimensional harmonic oscillator
\begin{gather*}
  E^{\rm osc}_N = 2(N + D), \qquad N = 2n_r + 2j = 0, 1, 2, \ldots,
\end{gather*}
whose levels, completely characterized by $N$, have a degeneracy equal to $\binom{N+2D-1}{2D-1}$.

Finally, the normalization constant in equation~(\ref{eq:wf-osc}) can be easily determined from some well-known properties of Laguerre and Jacobi polynomials~\cite{gradshteyn} and is given by
\begin{gather}
  {\cal N}^{\rm osc}_{n_r \nb \pb}  = \left(\frac{n_r!}{\pi^D (n_r+2j+D-1)!}\right)^{1/2} \nonumber\\
\phantom{{\cal N}^{\rm osc}_{n_r \nb \pb}  =}{}  \times \prod_{\nu=1}^{D-1} \left(\frac{n_{\nu}!  (2n_{\nu} + a_{\nu} + b_{\nu} + D-\nu-1) (n_{\nu}
       + a_{\nu} + b_{\nu} + D-\nu-2)!}{\left(n_{\nu} + a_{\nu} - \frac{1}{2}\right)!  \left(n_{\nu} + b_{\nu} + D
       - \nu - \frac{3}{2}\right)!}\right)^{1/2}.\!\!\!
   \label{eq:N-osc}
\end{gather}

\subsection{Harmonic oscillator symmetry and dynamical algebras}

As it is well known \cite{jauch, moshinsky}, to each of the oscillator levels specif\/ied by $N$ there corresponds a~symmetric unirrep $[N]$ of its su($2D$) symmetry algebra. The generators of the latter
\begin{gather*}
  \be_{\mu\nu} = E_{\mu\nu} - \frac{1}{2D} \delta_{\mu,\nu} \sum_{\rho} E_{\rho\rho}, \qquad \mu, \nu=1, 2,
  \ldots, 2D,
\end{gather*}
with
\begin{gather*}
  \left[\be_{\mu\nu}, \be_{\mu'\nu'}\right] = \delta_{\nu,\mu'} \be_{\mu\nu'} - \delta_{\mu,\nu'}
  \be_{\mu'\nu}, \qquad \be_{\mu\nu}^{\dagger} = \be_{\nu\mu},
\end{gather*}
are most easily constructed in terms of bosonic creation and annihilation operators
\begin{gather}
  \alpha_{\mu}^{\dagger} = \frac{1}{\sqrt{2}} \left(X_{\mu} - \partial_{X_{\mu}}\right), \qquad
  \alpha_{\mu} = \frac{1}{\sqrt{2}} \left(X_{\mu} + \partial_{X_{\mu}}\right), \qquad \mu=1, 2, \ldots, 2D,
  \label{eq:bosonic}
\end{gather}
from
\begin{gather}
  E_{\mu\nu} = \tfrac{1}{2} \{\alpha_{\mu}^{\dagger}, \alpha_{\nu}\} = \alpha_{\mu}^{\dagger}
  \alpha_{\nu} + \tfrac{1}{2} \delta_{\mu,\nu}.  \label{eq:unitary}
\end{gather}
The harmonic oscillator Hamiltonian turns out to be proportional to the f\/irst-order Casimir operator ${\cal C}_1$ of u($2D$),
\begin{gather}
  H^{\rm osc} = 2 {\cal C}_1 = 2 \sum_{\mu} E_{\mu\mu} = 2 \ce.  \label{eq:casimir}
\end{gather}

In the coordinates (\ref{eq:variables}) chosen to describe the oscillator, the so($2D$) subalgebra of su($2D$), generated by
\begin{gather}
  L_{\mu\nu} = - {\rm i} \left(\be_{\mu\nu} - \be_{\nu\mu}\right) = - {\rm i} (E_{\mu\nu} - E_{\nu\mu}),
  \label{eq:orthogonal}
\end{gather}
such that
\begin{gather*}
  [L_{\mu\nu}, L_{\mu'\nu'}] = {\rm i} (\delta_{\mu,\mu'} L_{\nu\nu'} - \delta_{\mu,\nu'} L_{\nu\mu'} -
  \delta_{\nu,\mu'} L_{\mu\nu'} + \delta_{\nu,\nu'} L_{\mu\mu'}), \qquad L_{\mu\nu}^{\dagger} =
  L_{\mu\nu} = - L_{\nu\mu},
\end{gather*}
is explicitly reduced. Its unirreps are characterized by $2j$, which runs over $N, N-2, \ldots, 0$ (or~1) for a given~$N$. The remaining generators of su($2D$) may be taken as
\begin{gather}
  T_{\mu\nu} = \be_{\mu\nu} + \be_{\nu\mu}.  \label{eq:T}
\end{gather}

The operators
\begin{gather}
  D^{\dagger}_{\mu\nu} = \alpha^{\dagger}_{\mu} \alpha^{\dagger}_{\nu}, \qquad D_{\mu\nu} =
  \alpha_{\mu} \alpha_{\nu}  \label{eq:D}
\end{gather}
act as raising and lowering operators relating among themselves wavefunctions corresponding to even or odd values of~$N$. Together with $E_{\mu\nu}$, they generate an ${\rm sp}(4D, \R)$ Lie algebra, whose (nonvanishing) commutation relations are given by
\begin{gather*}
    [E_{\mu\nu}, E_{\mu'\nu'}] = \delta_{\nu,\mu'} E_{\mu\nu'} - \delta_{\mu,\nu'} E_{\mu'\nu}, \\
   [E_{\mu\nu}, D^{\dagger}_{\mu'\nu'}] = \delta_{\nu,\mu'} D^{\dagger}_{\mu\nu'} + \delta_{\nu,\nu'}
           D^{\dagger}_{\mu\mu'}, \\
    [E_{\mu\nu}, D_{\mu'\nu'}] = - \delta_{\mu,\mu'} D_{\nu\nu'} - \delta_{\mu,\nu'} D_{\nu\mu'}, \\
    [D_{\mu\nu}, D^{\dagger}_{\mu'\nu'}] = \delta_{\mu,\mu'} E_{\nu'\nu} + \delta_{\mu,\nu'} E_{\mu'\nu}
          + \delta_{\nu,\mu'} E_{\nu'\mu} + \delta_{\nu,\nu'} E_{\mu'\mu}.
\end{gather*}

To connect the wavefunctions with an even $N$ value to those with an odd one, we have to use the bosonic creation and annihilation operators~(\ref{eq:bosonic}), which generate a Weyl algebra w($2D$), specif\/ied by
\begin{gather*}
  [\alpha_{\mu}, \alpha^{\dagger}_{\nu}] = \delta_{\mu,\nu} I.
\end{gather*}
The whole set of operators $\{E_{\mu\nu}, D^{\dagger}_{\mu\nu}, D_{\mu\nu}, \alpha^{\dagger}_{\mu}, \alpha_{\mu}, I\}$ then provides us with the harmonic oscillator dynamical algebra, which is the semidirect sum Lie algebra $\rm{w}(2D) \oplus_s {\rm sp}(4D, \R)$, as shown by the remaining (nonvanishing) commutation relations
\begin{alignat*}{3}
&  [E_{\mu\nu}, \alpha^{\dagger}_{\mu'}]  = \delta_{\nu,\mu'} \alpha^{\dagger}_{\mu}, \qquad &&
       [E_{\mu\nu}, \alpha_{\mu'}] = - \delta_{\mu,\mu'} \alpha_{\nu}, & \\
&  [D_{\mu\nu}, \alpha^{\dagger}_{\mu'}]  = \delta_{\mu,\mu'} \alpha_{\nu} + \delta_{\nu,\mu'} \alpha_{\mu},\qquad
       && [D^{\dagger}_{\mu\nu}, \alpha_{\mu'}] = - \delta_{\mu,\mu'} \alpha^{\dagger}_{\nu} -
       \delta_{\nu,\mu'} \alpha^{\dagger}_{\mu}.&
\end{alignat*}

To apply the symmetry and dynamical algebra generators to the oscillator wavefunctions~(\ref{eq:wf-osc}) written in the variables $R$, $\thetab$, $\lambdab$, we have to express the creation and annihilation operators~$\alpha^{\dagger}_{\mu}$,~$\alpha_{\mu}$ in such variables. This implies combining the transformation~(\ref{eq:variables}) with the corresponding change for the partial dif\/ferential operators
\begin{gather}
  \partial_{X_{2\nu-1}} = \sin \lambda_{\nu} \partial^{(\nu,1)} + \cos \lambda_{\nu} \partial^{(\nu,2)}, \nonumber\\
      \partial_{X_{2\nu}} = \cos \lambda_{\nu} \partial^{(\nu,1)} - \sin \lambda_{\nu} \partial^{(\nu,2)}, \qquad
  \nu=1, 2, \ldots, D,
   \label{eq:partial}
\end{gather}
where
\begin{gather*}
  \partial^{(1,1)}  = \sin \theta_1 \sin \theta_2 \cdots \sin \theta_{D-1} \partial_R \\
\phantom{\partial^{(1,1)}  =}{} + \frac{1}{R} \sum_{\rho=1}^{D-1} \csc \theta_1 \csc \theta_2 \cdots \csc \theta_{\rho-1}
        \cos \theta_{\rho} \sin \theta_{\rho+1} \sin \theta_{\rho+2} \cdots \sin \theta_{D-1}
        \partial_{\theta_{\rho}}, \\
  \partial^{(\nu,1)}  = \sin \theta_1 \sin \theta_2 \cdots \sin \theta_{D-\nu} \cos \theta_{D-\nu+1} \partial_R \\
\phantom{\partial^{(\nu,1)}  =}{} + \frac{1}{R} \sum_{\rho=1}^{D-\nu} \csc \theta_1 \csc \theta_2 \cdots \csc \theta_{\rho-1}
        \cos \theta_{\rho} \sin \theta_{\rho+1} \sin \theta_{\rho+2} \cdots \sin \theta_{D-\nu}
        \cos \theta_{D-\nu+1} \partial_{\theta_{\rho}} \\
\phantom{\partial^{(\nu,1)}  =}{}  - \frac{1}{R} \csc \theta_1 \csc \theta_2 \cdots \csc \theta_{D-\nu} \sin \theta_{D-\nu+1}
        \partial_{\theta_{D-\nu+1}}, \qquad \nu=2, 3, \ldots, D-1, \\
  \partial^{(D,1)}  = \cos \theta_1 \partial_R - \frac{1}{R} \sin \theta_1 \partial_{\theta_1},
\end{gather*}
and
\begin{gather*}
  \partial^{(1,2)}  = \frac{1}{R} \csc \theta_1 \csc \theta_2 \cdots \csc \theta_{D-1} \partial_{\lambda_1}, \\
  \partial^{(\nu,2)}  = \frac{1}{R} \csc \theta_1 \csc \theta_2 \cdots \csc \theta_{D-\nu} \sec \theta_{D-\nu+1}
        \partial_{\lambda_{\nu}}, \qquad \nu=2, 3, \ldots, D-1, \\
  \partial^{(D,2)}  = \frac{1}{R} \sec \theta_1 \partial_{\lambda_D}.
\end{gather*}
We shall carry out this transformation explicitly for $D=2$ in Section~\ref{section4}.

\section[Reduction of the $2D$-dimensional harmonic oscillator to the $D$-dimensional SW system]{Reduction of the $\boldsymbol{2D}$-dimensional harmonic oscillator\\ to the $\boldsymbol{D}$-dimensional SW system}\label{section3}

To go from the $2D$-dimensional harmonic oscillator Hamiltonian $H^{\rm osc}$ to some extended SW Hamiltonian $H$, let us f\/irst transform the original Cartesian coordinates $X_{\mu}$, $\mu=1, 2, \ldots,2D$, into some new ones $x_{\mu}$, $\mu=1, 2, \ldots,2D$, such that
\begin{gather*}
   x_1 = r \sin \phi_1 \sin \phi_2 \cdots \sin \phi_{D-1}, \\
   x_{\nu} = r \sin \phi_1 \sin \phi_2 \cdots \sin \phi_{D-\nu} \cos \phi_{D-\nu+1}, \qquad \nu=2, 3, \ldots,
        D-1, \\
  x_D = r \cos \phi_1, \\
   x_{D+\nu} = \lambda_{\nu}, \qquad \nu=1, 2, \ldots, D,
\end{gather*}
and
\begin{gather*}
 R = \sqrt{\omega}\, r, \qquad \theta_{\nu} = \phi_{\nu}, \quad \nu=1, 2, \ldots, D-1, \\
 0 \le r < \infty, \qquad 0 \le \phi_{\nu} < \frac{\pi}{2}, \quad \nu=1, 2, \ldots, D-1, \qquad 0 \le
       \lambda_{\nu} < 2\pi, \quad \nu=1, 2, \ldots, D.
\end{gather*}
Here $r$, $\phi_1, \phi_2, \ldots, \phi_{D-1}$ are hyperspherical coordinates in the $D$-dimensional subspace $(x_1, x_2$, $\ldots, x_D)$. The volume element in the transformed $2D$-dimensional space is given by
\begin{gather}
  dv = \prod_{\mu=1}^{2D} dx_{\mu} = r^{D-1} dr \left[\prod_{\nu=1}^{D-1} (\sin \phi_{\nu})^{D-\nu-1}
  d\phi_{\nu}\right] \left(\prod_{\nu=1}^D d\lambda_{\nu}\right).  \label{eq:dv}
\end{gather}

On making next the change of function
\begin{gather}
  \Psi(r, \phib, \lambdab) = {\cal O}^{1/2} \Psi^{\rm osc}(R, \thetab, \lambdab),  \label{eq:Psi}
\end{gather}
with
\begin{gather}
  {\cal O} = (\omega r)^D \prod_{\nu=1}^{D-1} (\sin \phi_{\nu})^{D-\nu} \cos \phi_{\nu},  \label{eq:O}
\end{gather}
the harmonic oscillator wavefunctions $\Psi^{\rm osc}(R, \thetab, \lambdab)$, living in a Hilbert space with measure~$dV$ given in~(\ref{eq:dV}), are mapped onto some functions $\Psi(r, \phib, \lambdab)$, living in a Hilbert space with measure~$dv$ def\/ined in~(\ref{eq:dv}). As a consequence of (\ref{eq:sp-osc}), we obtain
\begin{gather*}
  \int dv\, |\Psi(r, \phib, \lambdab)|^2 = 1.
\end{gather*}

By this unitary transformation, the harmonic oscillator Hamiltonian $H^{\rm osc}$ is changed into
\begin{gather}
  H/\omega = {\cal O}^{1/2} H^{\rm osc} {\cal O}^{-1/2}  \label{eq:H}
\end{gather}
and similarly for other operators acting in the harmonic oscillator Hilbert space. A straightforward calculation leads to the result
\begin{gather}
  H  = - \partial_r^2 - \frac{D-1}{r} \partial_r - \frac{1}{r^2} \Biggl\{\partial_{\phi_1}^2 + (D-2) \cot \phi_1
        \partial_{\phi_1} \nonumber\\
\phantom{H=}{} + \sum_{\nu=2}^{D-1} \frac{1}{\sin^2 \phi_1 \sin^2 \phi_2 \cdots \sin^2 \phi_{\nu-1}}
        \bigl[\partial_{\phi_{\nu}}^2 + (D-\nu-1) \cot \phi_{\nu} \partial_{\phi_{\nu}}\bigr] \nonumber\\
\phantom{H=}{}    + \frac{1}{\sin^2 \phi_1 \sin^2 \phi_2 \cdots \sin^2 \phi_{D-1}} \left(\partial_{\lambda_1}^2 +
        \frac{1}{4}\right) \nonumber\\
\phantom{H=}{} + \sum_{\nu=2}^{D-1} \frac{1}{\sin^2 \phi_1 \sin^2 \phi_2 \cdots \sin^2 \phi_{D-\nu}
        \cos^2 \phi_{D-\nu+1}} \left(\partial_{\lambda_{\nu}}^2 + \frac{1}{4}\right) + \frac{1}{\cos^2 \phi_1}
        \left(\partial_{\lambda_D}^2 + \frac{1}{4}\right)\Biggr\} \nonumber\\
\phantom{H=}{}           + \omega^2 r^2.\label{eq:H-bis}
\end{gather}

The eigenvalues of $H$ are directly obtained from (\ref{eq:E-osc}) as
\begin{gather}
  E_{n_r j} = 2\omega (2n_r + 2j + D), \qquad n_r=0, 1, 2, \ldots, \qquad j=0, \tfrac{1}{2}, 1, \tfrac{3}{2}, \ldots.
  \label{eq:E-H}
\end{gather}
The corresponding wavefunctions can be derived from (\ref{eq:wf-osc}), (\ref{eq:theta-bis}), (\ref{eq:theta}), (\ref{eq:L}), (\ref{eq:N-osc}), (\ref{eq:Psi}), and~(\ref{eq:O}) and read
\begin{gather}
  \Psi_{n_r\nb\pb}(r, \phib, \lambdab) = {\cal N}_{n_r\nb\pb} {\cal Z}^{(j)}_{n_r}(z) \Phi^{(\pb)}_{\nb}(\phib)
       \left(\prod_{\nu=1}^D e^{{\rm i} p_{D-\nu+1} \lambda_{\nu}}\right), \nonumber\\
  {\cal Z}^{(j)}_{n_r}(z) = \left(\frac{z}{\omega}\right)^{j + \frac{D}{4}} L_{n_r}^{(2j+D-1)}(z)
       e^{- \frac{1}{2} z}, \qquad z = \omega r^2,\nonumber \\
  \Phi^{(\pb)}_{\nb}(\phib) = \prod_{\nu=1}^{D-1} \Phi_{n_{\nu}}^{(a_{\nu}, b_{\nu})}(\phi_{\nu}) \nonumber\\
  \phantom{\Phi^{(\pb)}_{\nb}(\phib)}{}
    = \prod_{\nu=1}^{D-1} (\cos \phi_{\nu})^{a_{\nu}}
      (\sin \phi_{\nu})^{b_{\nu} + \frac{1}{2}(D-\nu-1)} P_{n_{\nu}}^{\left(a_{\nu} - \frac{1}{2}, b_{\nu} + D - \nu
      - \frac{3}{2}\right)}(- \cos 2\phi_{\nu}), \nonumber\\
   {\cal N}_{n_r\nb\pb} = \omega^{j + \frac{D}{2}} {\cal N}^{\rm osc}_{n_r\nb\pb},
  \label{eq:wf-H}
\end{gather}
where $\nb = n_1 n_2 \cdots n_{D-1}$, $\pb = p_1 p_2 \cdots p_D$, $n_r, n_1, n_2, \ldots, n_{D-1} \in \N$, $p_1, p_2, \ldots, p_D \in \Z$, while $j$, and $a_{\nu}$, $b_{\nu}$ are def\/ined in~(\ref{eq:j}) and~(\ref{eq:a-b}), respectively.

In the subspace of functions $\Psi_{n_r\nb\pb}(r, \phib, \lambdab)$ with f\/ixed $\pb$, the Hamiltonian $H$, def\/ined in~(\ref{eq:H-bis}), has the same action as the $D$-dimensional Hamiltonian
\begin{gather*}
  H^{(\kb)}  = - \partial_r^2 - \frac{D-1}{r} \partial_r - \frac{1}{r^2} \Biggl\{\partial_{\phi_1}^2 + (D-2) \cot \phi_1
        \partial_{\phi_1} \\
\phantom{H^{(\kb)}  =}{} + \sum_{\nu=2}^{D-1} \frac{1}{\sin^2 \phi_1 \sin^2 \phi_2 \cdots \sin^2 \phi_{\nu-1}}
        \bigl[\partial_{\phi_{\nu}}^2 + (D-\nu-1) \cot \phi_{\nu} \partial_{\phi_{\nu}}\bigr]\Biggr\} \\
\phantom{H^{(\kb)}  =}{} + \frac{k_1^2}{r^2 \sin^2 \phi_1 \sin^2 \phi_2 \cdots \sin^2 \phi_{D-1}} + \sum_{\nu=2}^{D-1}
        \frac{k_{\nu}^2}{r^2 \sin^2 \phi_1 \sin^2 \phi_2 \cdots \sin^2 \phi_{D-\nu} \cos^2 \phi_{D-\nu+1}} \\
 \phantom{H^{(\kb)}  =}{} + \frac{k_D^2}{r^2 \cos^2 \phi_1} + \omega^2 r^2,
\end{gather*}
where we have def\/ined $\kb = k_1 k_2 \cdots k_D$ and
\begin{gather}
  k_{\nu} = \sqrt{p_{D-\nu+1}^2 - \tfrac{1}{4}}, \qquad \nu=1, 2, \ldots, D.  \label{eq:k-p}
\end{gather}
The latter Hamiltonian is but the SW one (\ref{eq:H-SW}), expressed in hyperspherical coordinates $r$, $\phi_1, \phi_2, \ldots,\phi_{D-1}$. We conclude that $H$ is an extension of $H^{(\kb)}$, resulting from the introduction of $D$ auxiliary continuous variables $\lambda_{\nu} = x_{D+\nu}$, $\nu=1, 2, \ldots,D$, and that, conversely, $H^{(\kb)}$ is obtained from $H$ by projecting it down into the $D$-dimensional subspace $(x_1, x_2, \ldots, x_D)$.\footnote{Strictly speaking, this is true only for those~$k_{\nu}$'s that can be written in the form~(\ref{eq:k-p}) with integer $p_{D-\nu+1}^2$.\label{footnote3}}

As a by-product of this reduction process, we have determined the wavefunctions $\Psi^{(\kb)}_{n_r\nb}(r, \phib)$ of $H^{(\kb)}$ in hyperspherical coordinates. Equation (\ref{eq:wf-H}) may indeed be rewritten as
\begin{gather}
    \Psi_{n_r\nb\pb}(r, \phib, \lambdab) = \Psi^{(\kb)}_{n_r\nb}(r, \phib) (2\pi)^{-D/2} \prod_{\nu=1}^D
       e^{{\rm i} p_{D-\nu+1} \lambda_{\nu}}, \nonumber\\
    \Psi^{(\kb)}_{n_r\nb}(r, \phib) = {\cal N}^{(\kb)}_{n_r\nb} {\cal Z}^{(j)}_{n_r}(z) \Phi^{(\pb)}_{\nb}(\phib),
      \qquad {\cal N}^{(\kb)}_{n_r\nb} = (2\pi)^{D/2} {\cal N}_{n_r\nb\pb},
  \label{eq:wf-H-bis}
\end{gather}
with $\kb$ and $\pb$ related as in (\ref{eq:k-p}).

By a transformation similar to (\ref{eq:H}), the generators $\be_{\mu\nu}$ of the harmonic oscillator symmetry algebra su($2D$) are changed into some operators acting on $\Psi_{n_r\nb\pb}(r, \phib, \lambdab)$. Since the latter may change $n_r$ and $j$ separately (provided their sum $n_r + j = N/2$ is preserved), this means in particular (see equation (\ref{eq:j})) that the $p_{\nu}$'s (hence the $k_{\nu}$'s) may change too. The transformed su($2D$) algebra may therefore connect among themselves some wavefunctions of $H$ belonging to the same energy eigenvalue (\ref{eq:E-H}), but associated with dif\/ferent reduced Hamiltonians $H^{(\kb)}$. We conclude that it provides us with a potential algebra for the SW system. Similarly, the transformed ${\rm w}(2D) \oplus_s {\rm sp}(4D, \R)$ algebra will be a dynamical potential algebra for the same.

As a f\/inal point, it is worth observing that in the harmonic oscillator wavefunctions  (\ref{eq:wf-osc}), the quantum numbers $p_{\nu}$, $\nu=1, 2, \ldots,D$, run over $\Z$. On the other hand, in (\ref{eq:H-SW}), the parame\-ters~$k_{\nu}$, $\nu=1, 2, \ldots,D$, have been assumed real and positive. From equation~(\ref{eq:k-p}), however, it is clear that $p_{D-\nu+1} = 0$ would lead to an imaginary value of~$k_{\nu}$ and to unphysical wavefunctions (\ref{eq:wf-H-bis}), while $|p_{D-\nu+1}|$ and $- |p_{D-\nu+1}|$ with $|p_{D-\nu+1}| \ge 1$ would give rise to the same $k_{\nu}$, hence to some replicas of physical wavefunctions~(\ref{eq:wf-H-bis}). The correspondence between the harmonic oscillator wavefunctions and the extended SW ones is therefore not one-to-one. This lack of bijectiveness is a known aspect of potential algebraic approaches (see~\cite{cq88a} where this phenomenon was f\/irst pointed out).

\section{The two-dimensional case}\label{section4}

\subsection{Harmonic oscillator symmetry and dynamical algebras}\label{section4.1}

To deal in detail with the two-dimensional case, it is appropriate to rewrite the four-dimensional harmonic oscillator wavefunctions $\Psi^{\rm osc}_{n_r, n, p_1, p_2}(R, \theta, \lambda_1, \lambda_2)$ (with $j = n + \frac{1}{2}(|p_1| + |p_2|)$) in an equivalent form $\bpsi^{\rm osc}_{n_r, j, m, m'}(R, \theta, \lambda_1, \lambda_2)$ using either hyperspherical harmonics $Y_{2j, m, m'}(\alpha, \beta, \gamma)$ or (complex conjugate) rotation matrix elements $D^{j*}_{m, -m'}(\alpha, \beta, \gamma)$ expressed in terms of Euler angles $\alpha$, $\beta$, $\gamma$~\cite{biedenharn},
\begin{gather*}
    Y_{2j, m, m'}(\alpha, \beta, \gamma) = (-1)^{j-m'} \left(\frac{2j+1}{2\pi^2}\right)^{1/2}
       D^{j*}_{m, -m'}(\alpha, \beta, \gamma), \\
   D^{j*}_{m, -m'}(\alpha, \beta, \gamma) = e^{{\rm i}m\alpha} d^j_{m, -m'}(\beta) e^{- {\rm i}m'\gamma}.
\end{gather*}
Here $j$ runs over $0, \frac{1}{2}, 1, \frac{3}{2}, \ldots$, while $m$ and $m'$ take values in the set $\{j, j-1, \ldots, -j\}$. On setting
\begin{gather*}
  \theta = \tfrac{1}{2} \beta, \qquad \lambda_1 = \tfrac{1}{2}(\gamma - \alpha), \qquad \lambda_2 =
  \tfrac{1}{2}(\gamma + \alpha), \qquad p_1 = m - m', \qquad p_2 = - m - m',  
\end{gather*}
or, conversely,
\begin{gather*}
  \alpha = \lambda_2 - \lambda_1, \qquad \beta = 2\theta, \qquad \gamma = \lambda_2 + \lambda_1, \qquad
  m = \tfrac{1}{2}(p_1 - p_2), \qquad m' = - \tfrac{1}{2}(p_1 + p_2),
\end{gather*}
and on using the relation between rotation functions $d^j_{m, -m'}(\beta)$ and Jacobi polynomials \cite{biedenharn}, we indeed get
\begin{gather}
  \Psi^{\rm osc}_{n_r, n, p_1, p_2}(R, \theta, \lambda_1, \lambda_2) = (-1)^{\frac{1}{2}(|p_1| + p_1) + |p_2|}
  \bpsi^{\rm osc}_{n_r, j, m, m'}(R, \theta, \lambda_1, \lambda_2)  \label{eq:wf-osc-transf}
\end{gather}
with
\begin{gather}
   \bpsi^{\rm osc}_{n_r, j, m, m'}(R, \theta, \lambda_1, \lambda_2) = (-1)^{j-m'} \left(\frac{(2j+1) n_r!}{\pi^2
         (n_r + 2j + 1)!}\right)^{1/2} {\cal L}^{(j)}_{n_r}(z) \nonumber\\
\hphantom{\bpsi^{\rm osc}_{n_r, j, m, m'}(R, \theta, \lambda_1, \lambda_2) =}{}
\times d^j_{m,-m'}(2\theta) e^{- {\rm i}(m+m')\lambda_1} e^{{\rm i}(m-m')\lambda_2}, \nonumber\\
  {\cal L}^{(j)}_{n_r}(z) = z^j L^{(2j+1)}_{n_r}(z) e^{- \frac{1}{2} z}, \qquad z = R^2.
   \label{eq:wf-osc-bis}
\end{gather}

The advantage of this new form is that the wavefunctions $\Psi^{\rm osc}_{n_r, n, p_1, p_2}(R, \theta, \lambda_1, \lambda_2)$, which were classif\/ied according to
\begin{gather*}
\begin{matrix}
  {\rm su}(4) & \supset & {\rm so}(4) \\
  [N] &  & (2j)
\end{matrix}
\end{gather*}
with $N = 2n_r + 2j$ and $2j = 2n + |p_1| + |p_2|$, now turn out to be explicitly reduced with respect to
\begin{gather}
\begin{matrix}
  {\rm su}(4) & \supset & {\rm so}(4) \simeq {\rm su}(2) \oplus {\rm su}(2) & \supset & {\rm u}(1) \oplus
       {\rm u}(1) \\
  [N] & & (2j) \simeq [j] \oplus [j] & & [m] \oplus [m']
\end{matrix}.  \label{eq:chain}
\end{gather}
This will allow us to use the full machinery of angular momentum theory for determining the explicit action of the symmetry and dynamical algebra generators on wavefunctions.

The two su(2) algebras appearing in chain (\ref{eq:chain}) are generated by $J_i$ and $K_i$, $i=1, 2, 3$, def\/ined in terms of $L_{\mu\nu}$, $\mu, \nu=1, 2, 3, 4$, (see equation~(\ref{eq:orthogonal})) by
\begin{gather}
  J_i = \tfrac{1}{2}\left(\tfrac{1}{2} \epsilon_{ijk} L_{jk} - L_{i4}\right), \qquad K_i = \tfrac{1}{2}
  \left(\tfrac{1}{2}\epsilon_{ijk} L_{jk} + L_{i4}\right),  \label{eq:J-K}
\end{gather}
where $i$, $j$, $k$ run over 1, 2, 3 and $\epsilon_{ijk}$ is the antisymmetric tensor. The operators $J_i$ and $K_i$ satisfy the relations
\begin{gather*}
 [J_i, J_j] = {\rm i} \epsilon_{ijk} J_k, \qquad [K_i, K_j] = {\rm i} \epsilon_{ijk} K_k, \qquad [J_i, K_j] = 0, \\
  J_i^{\dagger} = J_i, \qquad K_i^{\dagger} = K_i.
\end{gather*}
Instead of the Cartesian components of $\mathbf J$ and $\mathbf K$, we may use alternatively $J_0 = J_3$, $J_{\pm} = J_1 \pm {\rm i} J_2$, $K_0 = K_3$, $K_{\pm} = K_1 \pm {\rm i} K_2$, with $J_0$ and $K_0$ generating the two u(1) subalgebras in (\ref{eq:chain}).

The dif\/ferential operator form of $J_0$, $J_{\pm}$, $K_0$, and $K_{\pm}$ can be obtained by combining equations (\ref{eq:variables}), (\ref{eq:bosonic}), (\ref{eq:unitary}), (\ref{eq:orthogonal}), (\ref{eq:partial}), and (\ref{eq:J-K}) and is given by
\begin{alignat*}{3}
&  J_0 = \tfrac{{\rm i}}{2} (\partial_{\lambda_1} - \partial_{\lambda_2}), \qquad && J_{\pm}  = \tfrac{1}{2}
      e^{\mp {\rm i}(\lambda_1 - \lambda_2)} [\pm \partial_{\theta} - {\rm i}(\cot \theta \partial_{\lambda_1}
      + \tan \theta \partial_{\lambda_2})], & \\
&  K_0 = \tfrac{{\rm i}}{2} (\partial_{\lambda_1} + \partial_{\lambda_2}), \qquad && K_{\pm}  = \tfrac{1}{2}
      e^{\mp {\rm i}(\lambda_1 + \lambda_2)} [\mp \partial_{\theta} + {\rm i}(\cot \theta \partial_{\lambda_1}
      - \tan \theta \partial_{\lambda_2})]. &
\end{alignat*}
From some dif\/ferential equation relations satisf\/ied by rotation functions $d^j_{m,-m'}(2\theta)$ \cite{schneider}, it is then easy to check that
\begin{alignat*}{3}
&  J_0 \bpsi^{\rm osc}_{n_r,j,m,m'} = m \bpsi^{\rm osc}_{n_r,j,m,m'}, \qquad && J_{\pm} \bpsi^{\rm osc}_{n_r,j,m,m'}
       = [(j\mp m)(j\pm m+1)]^{1/2} \bpsi^{\rm osc}_{n_r,j,m\pm1,m'}, & \\
 &  K_0 \bpsi^{\rm osc}_{n_r,j,m,m'}  = m' \bpsi^{\rm osc}_{n_r,j,m,m'}, \qquad && K_{\pm} \bpsi^{\rm osc}_{n_r,j,m,m'}
       = [(j\mp m')(j\pm m'+1)]^{1/2} \bpsi^{\rm osc}_{n_r,j,m,m'\pm1},&
\end{alignat*}
which proves the above-mentioned result.

It is now convenient to rewrite all operators of physical interest as components $T^{(s,t)}_{\sigma, \tau}$, $\sigma=s$, $s-1, \ldots,-s$, $\tau=t, t-1, \ldots,-t$, of irreducible tensors of rank $(s,t)$ with respect to ${\rm su}(2) \oplus {\rm su}(2)$. These must satisfy commutation relations of the type
\begin{alignat*}{3}
 & \big[J_0, T^{(s,t)}_{\sigma, \tau}\big]  = \sigma T^{(s,t)}_{\sigma, \tau},\qquad
       && \big[J_{\pm}, T^{(s,t)}_{\sigma, \tau}\big] = [(s\mp\sigma) (s\pm\sigma+1)]^{1/2}
       T^{(s,t)}_{\sigma\pm1, \tau}, & \\
&  \big[K_0, T^{(s,t)}_{\sigma, \tau}\big] = \tau T^{(s,t)}_{\sigma, \tau},\qquad
       &&  \big[K_{\pm}, T^{(s,t)}_{\sigma, \tau}\big] = [(t\mp\tau) (t\pm\tau+1)]^{1/2}
       T^{(s,t)}_{\sigma, \tau\pm1}.&
\end{alignat*}

Since the bosonic creation and annihilation operators serve as building blocks for the construction of other operators, let us start with them. The creation operators can be written as components $\ca^{\dagger}_{\sigma, \tau}$, $\sigma$, $\tau = \frac{1}{2}$, $- \frac{1}{2}$, of an irreducible tensor of rank $\left(\frac{1}{2}, \frac{1}{2}\right)$,
\begin{gather}
  \ca^{\dagger}_{\pm\frac{1}{2}, \pm\frac{1}{2}} = \mp \tfrac{1}{\sqrt{2}} \big(\alpha^{\dagger}_1 \pm {\rm i}
  \alpha^{\dagger}_2\big), \qquad \ca^{\dagger}_{\pm\frac{1}{2}, \mp\frac{1}{2}} = \tfrac{1}{\sqrt{2}}
  \big(\alpha^{\dagger}_3 \mp {\rm i} \alpha^{\dagger}_4\big).  \label{eq:bosonic-tensor}
\end{gather}
The same is true for the annihilation operators, the corresponding components being given by
\begin{gather}
  \ca_{\sigma, \tau} = (-1)^{1-\sigma-\tau} \big(\ca^{\dagger}_{-\sigma, -\tau}\big)^{\dagger}, \qquad
  \sigma, \tau = \tfrac{1}{2}, - \tfrac{1}{2}.  \label{eq:annihilation}
\end{gather}

On coupling an operator $\ca^{\dagger}$ with an operator $\ca$ according to
\begin{gather*}
  \big[\ca^{\dagger} \times \ca\big]^{s,t}_{\sigma,\tau} = \sum_{\sigma',\tau'} \bigl\langle \tfrac{1}{2}\,
  \sigma', \tfrac{1}{2}\, \sigma-\sigma' \big| s\, \sigma\bigr\rangle \bigl\langle \tfrac{1}{2}\, \tau',
  \tfrac{1}{2}\, \tau-\tau' \big| t\, \tau\bigr\rangle \ca^{\dagger}_{\sigma',\tau'}
  \ca_{\sigma-\sigma',\tau-\tau'},
\end{gather*}
where $\langle\, , \, | \,\rangle$ denotes an su(2) Wigner coef\/f\/icient \cite{biedenharn}, we obtain the su(4) symmetry algebra generators classif\/ied with respect to chain (\ref{eq:chain}). These include
\begin{gather*}
  J_{\sigma} = \big[\ca^{\dagger} \times \ca\big]^{1,0}_{\sigma,0}, \qquad K_{\tau} = \big[\ca^{\dagger}
  \times \ca\big]^{0,1}_{0,\tau}, \qquad \sigma, \tau=+1, 0, -1,
\end{gather*}
with $J_{\pm1} = \mp J_{\pm}/\sqrt{2}$ and $K_{\pm1} = \mp K_{\pm}/\sqrt{2}$, as well as the nine components of an irreducible tensor of rank $(1,1)$,
\begin{gather}
  \ct_{\sigma,\tau} = \big[\ca^{\dagger} \times \ca\big]^{1,1}_{\sigma,\tau}, \qquad \sigma, \tau = +1, 0,
  -1.  \label{eq:T-tensor}
\end{gather}
The latter may be written as
\begin{alignat*}{3}
& \ct_{\pm1, \pm1} = - \tfrac{1}{4}(T_{11} \pm 2{\rm i} T_{12} - T_{22}), \qquad && \ct_{\pm1, 0} =
        \tfrac{1}{2\sqrt{2}}(\pm T_{13} - {\rm i}T_{14} + {\rm i}T_{23} \pm T_{24}), & \\
  & \ct_{\pm1, \mp1} = - \tfrac{1}{4}(T_{33} \mp 2{\rm i} T_{34} - T_{44}), \qquad && \ct_{0, \pm1} =
        \tfrac{1}{2\sqrt{2}}(\pm T_{13} + {\rm i}T_{14} + {\rm i}T_{23} \mp T_{24}),&  \\
  & \ct_{0,0} = \tfrac{1}{2} (T_{11} + T_{22}) = - \tfrac{1}{2}(T_{33} + T_{44}) &&&
\end{alignat*}
in terms of the operators $T_{\mu\nu}$, def\/ined in (\ref{eq:T}). Observe that the u(4) f\/irst-order Casimir operator (\ref{eq:casimir}) is, up to some constants, the scalar that can be obtained in such a coupling procedure,
\begin{gather*}
  \big[\ca^{\dagger} \times \ca\big]^{0,0}_{0,0} = \tfrac{1}{2}(\ce - 2).
\end{gather*}

Similarly, the coupling of two operators $\ca^{\dagger}$ provides us with the raising operators belonging to ${\rm sp}(8, \R)$,
\begin{gather*}
    \cd^{\dagger} = \ca^{\dagger} \cdot \ca^{\dagger} = - 2 \big[\ca^{\dagger} \times \ca^{\dagger}\big]
       ^{0,0}_{0,0}, \qquad
   \cd^{\dagger}_{\sigma,\tau} = \big[\ca^{\dagger} \times \ca^{\dagger}\big]^{1,1}_{\sigma,\tau}, \qquad
       \sigma, \tau = +1, 0, -1,
\end{gather*}
or, in detail,
\begin{gather*}
  \cd^{\dagger} = D^{\dagger}_{11} + D^{\dagger}_{22} + D^{\dagger}_{33} + D^{\dagger}_{44}
\end{gather*}
and
\begin{alignat*}{3}
  & \cd^{\dagger}_{\pm1, \pm1} = \frac{1}{2}\bigl(D^{\dagger}_{11} \pm 2{\rm i} D^{\dagger}_{12} -
        D^{\dagger}_{22}\bigr), \qquad &&  \cd^{\dagger}_{\pm1, 0} = - \frac{1}{\sqrt{2}}\bigl(\pm
        D^{\dagger}_{13} - {\rm i} D^{\dagger}_{14} + {\rm i}D^{\dagger}_{23} \pm D^{\dagger}_{24}\bigr), & \\
  & \cd^{\dagger}_{\pm1, \mp1} = \frac{1}{2}\bigl(D^{\dagger}_{33} \mp 2{\rm i} D^{\dagger}_{34} -
        D^{\dagger}_{44}\bigr), \qquad && \cd^{\dagger}_{0, \pm1} = - \frac{1}{\sqrt{2}}\bigl(\pm
        D^{\dagger}_{13} + {\rm i}D^{\dagger}_{14} + {\rm i}D^{\dagger}_{23} \mp D^{\dagger}_{24}\bigr), & \\
  & \cd^{\dagger}_{0,0} = \frac{1}{2} \bigl(- D^{\dagger}_{11} - D^{\dagger}_{22} + D^{\dagger}_{33} +
        D^{\dagger}_{44}\bigr) &&&
\end{alignat*}
in terms of $D^{\dagger}_{\mu\nu}$ def\/ined in (\ref{eq:D}). The corresponding lowering operators are then
\begin{gather*}
  \cd = \bigl(\cd^{\dagger}\bigr)^{\dagger}, \qquad \cd_{\sigma,\tau} = (-1)^{\sigma+\tau}
  \bigl(\cd^{\dagger}_{-\sigma,-\tau}\bigr)^{\dagger}, \qquad \sigma, \tau=+1, 0, -1.
\end{gather*}

It is now straightforward to determine the action of $\ca^{\dagger}_{\sigma,\tau}$ on the wavefunctions $\bpsi^{\rm osc}_{n_r, j, m, m'}(R, \theta$, $\lambda_1, \lambda_2)$. Application of the Wigner--Eckart theorem with respect to ${\rm su}(2) \oplus {\rm su}(2)$ \cite{biedenharn} indeed leads to the relation
\begin{gather}
  \ca^{\dagger}_{\sigma,\tau} \bpsi^{\rm osc}_{n_r, j, m, m'}  = \sum_{n'_r,j'} \bigl\langle n'_r, j' \big\|
      \ca^{\dagger} \big\| n_r, j \bigr\rangle \bigl\langle j\, m, \tfrac{1}{2}\, \sigma \big| j'\, m+\sigma
      \bigr\rangle \bigl\langle j\, m', \tfrac{1}{2}\, \tau \big| j'\, m'+\tau \bigr\rangle \nonumber\\
\phantom{\ca^{\dagger}_{\sigma,\tau} \bpsi^{\rm osc}_{n_r, j, m, m'}  =}{}  \times \bpsi^{\rm osc}_{n'_r, j', m+\sigma, m'+\tau},
  \label{eq:creation-action}
\end{gather}
where $\bigl\langle n'_r, j' \big\| \ca^{\dagger} \big\| n_r, j \bigr\rangle$ denotes a reduced matrix element, the summation over $j'$ runs over $j + \frac{1}{2}, j - \frac{1}{2}$, and $n'_r$ is determined by the selection rule $n'_r + j' = n_r + j + \frac{1}{2}$ implying that $n'_r = n_r, n_r+1$, respectively. To calculate the two independent reduced matrix elements, it is enough to consider equation (\ref{eq:creation-action}) for the special case $m = m' = j$ and to use the dif\/ferential operator form of $\ca^{\dagger}_{\pm \frac{1}{2}, \pm \frac{1}{2}}$,
\begin{gather*}
  \ca^{\dagger}_{\pm \frac{1}{2}, \pm \frac{1}{2}} = \frac{1}{2} e^{\mp {\rm i}\lambda_1} \left[{\rm i}
  \left(\sin \theta \partial_R + \frac{1}{R} \cos \theta \partial_{\theta}\right) \pm \frac{1}{R} \csc \theta
  \partial_{\lambda_1} - {\rm i} R \sin \theta\right],
\end{gather*}
following from (\ref{eq:variables}), (\ref{eq:bosonic}), (\ref{eq:partial}), and (\ref{eq:bosonic-tensor}). Simple properties of the rotation function $d^j_{m,-m'}(2\theta)$ and of the Laguerre polynomial $L^{(2j+1)}_{n_r}(z)$ then lead to the results
\begin{gather}
   \bigl\langle n_r, j + \tfrac{1}{2} \big\| \ca^{\dagger} \big\| n_r, j \bigr\rangle = {\rm i} \left(\frac{(2j+1)
       (n_r+2j+2)}{2j+2}\right)^{1/2}, \nonumber\\
  \bigl\langle n_r + 1, j - \tfrac{1}{2} \big\| \ca^{\dagger} \big\| n_r, j \bigr\rangle = - {\rm i} \left(\frac{(2j+1)
       (n_r+1)}{2j}\right)^{1/2}.
  \label{eq:creation-red}
\end{gather}

The operators $\ca_{\sigma,\tau}$ satisfy an equation similar to (\ref{eq:creation-action}) with $\bigl\langle n'_r, j' \big\| \ca^{\dagger} \big\| n_r, j \bigr\rangle$ replaced by $\bigl\langle n'_r, j' \big\| \ca \big\| n_r, j \bigr\rangle$ and $n'_r = n_r - 1, n_r$ for $j' = j + \frac{1}{2}, j - \tfrac{1}{2}$, respectively. The corresponding reduced matrix elements can be directly calculated from the relation
\begin{gather}
  \bigl\langle n'_r, j' \big\| \ca \big\| n_r, j \bigr\rangle = \frac{2j+1}{2j'+1} \bigl\langle n_r, j \big\|
  \ca^{\dagger} \big\| n'_r, j' \bigr\rangle^*,  \label{eq:annihilation-red}
\end{gather}
which is a direct consequence of (\ref{eq:annihilation}).

For the su(4) generators that do not belong to so(4), we get the equation
\begin{gather}
  \ct_{\sigma,\tau} \bpsi^{\rm osc}_{n_r, j, m, m'}  = \sum_{n'_r,j'} \bigl\langle n'_r, j' \big\|
      \ct \big\| n_r, j \bigr\rangle \bigl\langle j\, m, 1\, \sigma \big| j'\, m+\sigma \bigr\rangle
      \bigl\langle j\, m', 1\, \tau \big| j'\, m'+\tau \bigr\rangle \nonumber\\
\phantom{\ct_{\sigma,\tau} \bpsi^{\rm osc}_{n_r, j, m, m'}  =}{}  \times \bpsi^{\rm osc}_{n'_r, j', m+\sigma, m'+\tau},
   \label{eq:T-action}
\end{gather}
where $j' = j+1, j, j-1$ and $n'_r = n_r-1, n_r, n_r+1$, respectively. Equation (\ref{eq:T-tensor}) and the coupling law for reduced matrix elements~\cite{biedenharn} enable us to determine
\begin{gather*}
   \bigl\langle n_r - 1, j + 1 \big\| \ct \big\| n_r, j \bigr\rangle = - \left(\frac{(2j+1)n_r(n_r+2j+2)}{2j+3}
       \right)^{1/2}, \\
    \bigl\langle n_r, j \big\| \ct \big\| n_r, j \bigr\rangle = n_r + j +1, \\
   \bigl\langle n_r + 1, j - 1 \big\| \ct \big\| n_r, j \bigr\rangle = - \left(\frac{(2j+1)(n_r+1)(n_r+2j+1}{2j-1}
  \right)^{1/2}
\end{gather*}
from (\ref{eq:creation-red}) and (\ref{eq:annihilation-red}).

The operators $\cd^{\dagger}_{\sigma,\tau}$ and $\cd_{\sigma,\tau}$ satisfy a relation similar to (\ref{eq:T-action}) with
\begin{gather*}
    \bigl\langle n_r, j + 1 \big\| \cd^{\dagger} \big\| n_r, j \bigr\rangle = - \left(
       \frac{(2j+1)(n_r+2j+2)(n_r+2j+3)}{2j+3}\right)^{1/2}, \\
    \bigl\langle n_r + 1, j \big\| \cd^{\dagger}\big\| n_r, j \bigr\rangle = [(n_r+1)(n_r + 2j + 2)]^{1/2}, \\
    \bigl\langle n_r + 2, j - 1 \big\| \cd^{\dagger} \big\| n_r, j \bigr\rangle = - \left(
       \frac{(2j+1)(n_r+1)(n_r+2)}{2j-1}\right)^{1/2},
\end{gather*}
and $\bigl\langle n'_r, j' \big\| \cd\big\| n_r, j \bigr\rangle$ obtained from these as in (\ref{eq:annihilation-red}).

Finally, with the equations
\begin{gather*}
 \cd^{\dagger} \bpsi^{\rm osc}_{n_r, j, m, m'} = -2 [(n_r+1)(n_r+2j+2)]^{1/2}
      \bpsi^{\rm osc}_{n_r+1, j, m, m'}, \\
 \cd \bpsi^{\rm osc}_{n_r, j, m, m'} = -2 [n_r(n_r+2j+1)]^{1/2} \bpsi^{\rm osc}_{n_r-1, j, m, m'},
\end{gather*}
the action of the harmonic oscillator symmetry and dynamical algebra generators on $\bpsi^{\rm osc}_{n_r, j, m, m'}\!(\!R$, $\theta, \lambda_1, \lambda_2)$ is completely determined.

\subsection{SW system potential and dynamical potential algebras}

In two dimensions, equations (\ref{eq:a-b}) and (\ref{eq:k-p}) simply lead to
\begin{gather*}
  k_1^2 = b(b-1), \qquad k_2^2 = a(a-1).
\end{gather*}
In the following, it will prove convenient to use $a$ and $b$ instead of $k_1$ and $k_2$. Up to the same phase factor as that occurring in (\ref{eq:wf-osc-transf}), the extended SW Hamiltonian wavefunctions (\ref{eq:wf-H-bis}) can then be rewritten as\footnote{It is worth observing here that integer or half-integer values of~$j$, $m$, and $m'$ are related to integer values of~$n$ and half-integer ones of $a$ and $b$. The results for matrix elements of potential and dynamical potential algebra generators are only valid for such~$a$ and~$b$ (see footnote~\ref{footnote3}), although those for wavefunctions are not restricted to these values provided factorials are replaced by gamma functions.}
\begin{gather*}
   \bpsi_{n_r,n,a,b}(r,\phi,\lambda_1,\lambda_2) = \bpsi^{(a,b)}_{n_r,n}(r,\phi) (2\pi)^{-1}
      e^{{\rm i} \left(b-\frac{1}{2}\right) \lambda_1} e^{{\rm i} \left(a-\frac{1}{2}\right) \lambda_2}, \\
   \bpsi^{(a,b)}_{n_r,n}(r,\phi) = {\cal N}^{(a,b)}_{n_r,n} {\cal Z}^{(j)}_{n_r}(z) \Phi^{(a,b)}_n(\phi), \qquad
   {\cal Z}^{(j)}_{n_r}(z) = \left(\frac{z}{\omega}\right)^{n + \frac{1}{2}(a+b)} L^{(2n+a+b)}_{n_r}(z)
      e^{- \frac{1}{2} z}, \\
   \Phi^{(a,b)}_n(\phi) = \cos^a \phi \sin^b \phi P^{\left(a-\frac{1}{2}, b-\frac{1}{2}\right)}_n(- \cos 2\phi), \\
   {\cal N}^{(a,b)}_{n_r,n} = (-1)^{a+b-1} 2 \left(\frac{\omega^{2n+a+b+1} n_r!\, n!\, (2n+a+b) (n+a+b-1)!}
      {(n_r+2n+a+b)! \left(n+a-\frac{1}{2}\right)! \left(n+b-\frac{1}{2}\right)!}\right)^{1/2},
\end{gather*}
where $n$, $a$, $b$ are related to $j$, $m$, $m'$ used in (\ref{eq:wf-osc-bis}) through the relations\footnote{Equation (\ref{eq:j-m-m'}) is valid for positive $p_1$ and $p_2$, corresponding to physical wavefunctions (see discussion at the end of Section~\ref{section3}).}
\begin{gather}
  j = n + \tfrac{1}{2}(a+b-1), \qquad m = \tfrac{1}{2}(a-b), \qquad m' = - \tfrac{1}{2}(a+b-1),
  \label{eq:j-m-m'}
\end{gather}
or, conversely,
\begin{gather*}
  a = m-m'+\tfrac{1}{2}, \qquad b = -m-m'+\tfrac{1}{2}, \qquad n = j+m'.
\end{gather*}

The generators of the su(4) potential algebra, as well as those of the ${\rm w}(4) \oplus_s {\rm sp}(8, \R)$ dynamical potential algebra, can be directly obtained by performing transformation (\ref{eq:H}) on the operators of Section~\ref{section4.1}. We get for instance\footnote{For simplicity's sake, we denote both types of operators by the same symbols.}
\begin{gather*}
  J_0 = \frac{{\rm i}}{2} (\partial_{\lambda_1} - \partial_{\lambda_2}), \qquad K_0 = \frac{{\rm i}}{2}
  (\partial_{\lambda_1} + \partial_{\lambda_2}),
\\
  J_{\pm} = \frac{1}{2} e^{\mp{\rm i}(\lambda_1-\lambda_2)} \left[\pm \partial_{\phi} - \cot \phi \left({\rm i}
  \partial_{\lambda_1} \pm \frac{1}{2}\right) - \tan \phi \left({\rm i} \partial_{\lambda_2} \mp \frac{1}{2}\right)
  \right],
\\
  K_{\pm} = \frac{1}{2} e^{\mp{\rm i}(\lambda_1+\lambda_2)} \left[\mp \partial_{\phi} + \cot \phi \left({\rm i}
  \partial_{\lambda_1} \pm \frac{1}{2}\right) - \tan \phi \left({\rm i} \partial_{\lambda_2} \pm \frac{1}{2}\right)
  \right],
\\
  \ct_{+1,+1}  = \frac{1}{4\omega} e^{-2{\rm i}\lambda_1} \biggl[- \sin^2 \phi \partial^2_r - \frac{2}{r} \sin \phi
      \cos \phi \partial^2_{r\phi} + \frac{{2\rm i}}{r} \partial^2_{r\lambda_1} - \frac{1}{r^2} \cos^2 \phi
      \partial^2_{\phi} \\
\phantom{\ct_{+1,+1}  =}{}  + \frac{2{\rm i}}{r^2} \cot \phi \partial^2_{\phi\lambda_1} + \frac{1}{r^2} \csc^2 \phi
      \partial^2_{\lambda_1} + \frac{1}{r}(1 + \sin^2\phi) \partial_r + \frac{2}{r^2} (\cot\phi + \sin\phi \cos\phi)
      \partial_{\phi} \\
\phantom{\ct_{+1,+1}  =}{} - \frac{3{\rm i}}{r^2} \csc^2\phi \partial_{\lambda_1} - \frac{5}{4r^2} \csc^2\phi + \omega^2 r^2
      \sin^2\phi\biggr],
\\
  \ca^{\dagger}_{\pm\frac{1}{2}, \pm\frac{1}{2}} = \frac{1}{2} e^{\mp{\rm i}\lambda_1} \left[{\rm i} \left(
  \sin\phi \partial_r + \frac{1}{r} \cos\phi \partial_{\phi}\right) \pm \frac{1}{r} \csc\phi \partial_{\lambda_1}
  - \frac{{\rm i}}{2r} \csc\phi - {\rm i}r \sin\phi\right],
\\
  \cd^{\dagger}_{+1,+1}  = \frac{1}{4\omega} e^{-2{\rm i}\lambda_1} \biggl\{- \sin^2 \phi \partial^2_r -
      \frac{2}{r} \sin\phi \cos \phi \partial^2_{r\phi} + \frac{{2\rm i}}{r} \partial^2_{r\lambda_1} - \frac{1}{r^2}
      \cos^2 \phi \partial^2_{\phi} + \frac{2{\rm i}}{r^2} \cot \phi \partial^2_{\phi\lambda_1} \\
\phantom{\cd^{\dagger}_{+1,+1}  =}{} + \frac{1}{r^2} \csc^2 \phi \partial^2_{\lambda_1} + \biggl[\frac{1}{r}(1 + \sin^2\phi) + 2\omega r
      \sin^2\phi\biggr] \partial_r + 2 \biggl[\frac{1}{r^2} (\cot\phi + \sin\phi \cos\phi) \\
\phantom{\cd^{\dagger}_{+1,+1}  =}{}    + \omega \sin\phi \cos\phi\biggr] \partial_{\phi} - {\rm i} \biggl(\frac{3}{r^2} \csc^2\phi + 2\omega
      \biggr) \partial_{\lambda_1} - \frac{5}{4r^2} \csc^2\phi - \omega^2 r^2 \sin^2\phi - \omega\biggr\},
\\
  \cd^{\dagger} = \frac{1}{2\omega} \bigl(- H - 2\omega r \partial_r + 2 \omega^2 r^2 - 2\omega\bigr).
\end{gather*}

It is also straightforward to derive their matrix elements from the results of Section~\ref{section4.1} and equation (\ref{eq:j-m-m'}). We list them below:
\begin{gather*}
  J_0 \bpsi_{n_r,n,a,b} = \tfrac{1}{2}(a-b) \bpsi_{n_r,n,a,b}, \qquad K_0 \bpsi_{n_r,n,a,b} = - \tfrac{1}{2}(a+b-1)
  \bpsi_{n_r,n,a,b},
\\
  J_+ \bpsi_{n_r,n,a,b} = \left[\left(n+a+\tfrac{1}{2}\right) \left(n+b-\tfrac{1}{2}\right)\right]^{1/2}
  \bpsi_{n_r,n,a+1,b-1},
\\
  J_- \bpsi_{n_r,n,a,b} = \left[\left(n+a-\tfrac{1}{2}\right) \left(n+b+\tfrac{1}{2}\right)\right]^{1/2}
  \bpsi_{n_r,n,a-1,b+1},
\\
  K_+ \bpsi_{n_r,n,a,b} = [(n+1) (n+a+b-1)]^{1/2} \bpsi_{n_r,n+1,a-1,b-1},
\\
  K_- \bpsi_{n_r,n,a,b} = [n (n+a+b)]^{1/2} \bpsi_{n_r,n-1,a+1,b+1},
\\
  \ct_{\sigma,\tau} \bpsi_{n_r,n,a,b} = \sum_{n'=n+\tau-1}^{n+\tau+1} t_{n'}(n_r, 2n+a+b) \\
   \quad {}\times \bigl\langle n + \tfrac{1}{2}(a+b-1)\: \tfrac{1}{2}(a-b), 1\: \sigma \big| n' - \tau +
       \tfrac{1}{2}(a+b-1)\: \tfrac{1}{2}(a-b) + \sigma \bigr\rangle \\
   \quad {}\times \bigl\langle n + \tfrac{1}{2}(a+b-1)\: - \tfrac{1}{2}(a+b-1), 1\: \tau \big| n' - \tau +
       \tfrac{1}{2}(a+b-1)\: - \tfrac{1}{2}(a+b-1) + \tau \bigr\rangle \\
 \quad {}\times \bpsi_{n_r-(n'-n-\tau), n', a+\sigma-\tau, b-\sigma-\tau},
\\
    \ca^{\dagger}_{\sigma,\tau} \bpsi_{n_r,n,a,b} = \sum_{n'=n+\tau-\frac{1}{2}}^{n+\tau+\frac{1}{2}}
        a_{n'}(n_r, 2n+a+b) \\
   \quad {}\times \bigl\langle n + \tfrac{1}{2}(a+b-1)\: \tfrac{1}{2}(a-b), \tfrac{1}{2}\: \sigma \big| n' - \tau +
       \tfrac{1}{2}(a+b-1)\: \tfrac{1}{2}(a-b) + \sigma \bigr\rangle \\
   \quad {}\times \bigl\langle n + \tfrac{1}{2}(a+b-1)\: - \tfrac{1}{2}(a+b-1), \tfrac{1}{2}\: \tau \big| n' - \tau +
       \tfrac{1}{2}(a+b-1)\: - \tfrac{1}{2}(a+b-1) + \tau \bigr\rangle \\
   \quad {}\times \bpsi_{n_r-(n'-n-\tau)+\frac{1}{2}, n', a+\sigma-\tau, b-\sigma-\tau},
\\
    \cd^{\dagger}_{\sigma,\tau} \bpsi_{n_r,n,a,b} = \sum_{n'=n+\tau-1}^{n+\tau+1} d_{n'}(n_r, 2n+a+b) \\
   \quad {}\times \bigl\langle n + \tfrac{1}{2}(a+b-1)\: \tfrac{1}{2}(a-b), 1\: \sigma \big| n' - \tau +
       \tfrac{1}{2}(a+b-1)\: \tfrac{1}{2}(a-b) + \sigma \bigr\rangle \\
   \quad {}\times \bigl\langle n + \tfrac{1}{2}(a+b-1)\: - \tfrac{1}{2}(a+b-1), 1\: \tau \big| n' - \tau +
       \tfrac{1}{2}(a+b-1)\: - \tfrac{1}{2}(a+b-1) + \tau \bigr\rangle \\
   \quad {}\times \bpsi_{n_r-(n'-n-\tau)+1, n', a+\sigma-\tau, b-\sigma-\tau},
\\
  \cd^{\dagger} \bpsi_{n_r,n,a,b} = - 2 [(n_r+1) (n_r+2n+a+b+1)]^{1/2} \bpsi_{n_r+1,n,a,b}.
\end{gather*}
Here
\begin{gather*}
  t_{n'}(n_r, 2n+a+b) =
  \begin{cases}
    - \left(\frac{(2n+a+b)n_r(n_r+2n+a+b+1)}{2n+a+b+2}\right)^{1/2}& \text{if $n' = n+\tau+1$}, \\
    n_r + n + \tfrac{1}{2}(a+b+1)& \text{if $n' = n+\tau$}, \\
    - \left(\frac{(2n+a+b)(n_r+1)(n_r+2n+a+b)}{2n+a+b-2}\right)^{1/2}& \text{if $n' = n+\tau-1$},
  \end{cases}
\\
  a_{n'}(n_r, 2n+a+b) =
  \begin{cases}
    {\rm i} \left(\frac{(2n+a+b)(n_r+2n+a+b+1)}{2n+a+b+1}\right)^{1/2}& \text{if $n' = n+\tau+\tfrac{1}{2}$}, \\
    - {\rm i} \left(\frac{(2n+a+b)(n_r+1)}{2n+a+b-1}\right)^{1/2}& \text{if $n' = n+\tau-\tfrac{1}{2}$},
  \end{cases}
\end{gather*}
and
\begin{gather*}
  d_{n'}(n_r, 2n+a+b) =
  \begin{cases}
    - \left(\frac{(2n+a+b)(n_r+2n+a+b+1)(n_r+2n+a+b+2)}{2n+a+b+2}\right)^{1/2}& \text{if $n' = n+\tau+1$}, \\
    [(n_r + 1)(n_r + 2n + a+b+1)]^{1/2}& \text{if $n' = n+\tau$}, \\
    - \left(\frac{(2n+a+b)(n_r+1)(n_r+2)}{2n+a+b-2}\right)^{1/2}& \text{if $n' = n+\tau-1$}.
  \end{cases}
\end{gather*}

From these results, we conclude that the potential algebra generators produce transitions between levels belonging to spectra of Hamiltonians characterized by parameters $(a, b)$, $(a\pm1$, $b\mp1)$, $(a\pm1, b\pm1)$, $(a\pm2, b)$, and $(a, b\pm2)$. For the dynamical potential algebra generators, the same Hamiltonians are involved together with those associated with $(a\pm1, b)$ and $(a, b\pm1)$.

\section{Conclusion}\label{section5}

In the present paper, we have re-examined the $D$-dimensional SW system, which may be considered as the archetype of $D$-dimensional superintegrable system. We have completed Evans previous algebraic study, wherein its symmetry and dynamical algebras had been determined, by constructing its potential and dynamical potential algebras.

In our approach based on the use of hyperspherical coordinates in the $D$-dimensional space and on the introduction of $D$ auxiliary continuous variables, the SW system has been obtained by reducing a $2D$-dimensional harmonic oscillator Hamiltonian. The su($2D$) symmetry and ${\rm w}(2D) \oplus_s {\rm sp}(4D, \R)$ dynamical algebras of the latter have then been transformed into correspon\-ding potential and dynamical potential algebras for the former. Finally, the two-dimensional case has been studied in the fullest detail.

Possible connections with other approaches currently used in connection with the SW system or, more generally, superintegrable systems, such as supersymmetry \cite{marquette09c, marquette10}, path integrals \cite{grosche}, coherent states \cite{unal}, and deformations \cite{herranz, carinena}, might be interesting topics for future investigation.

\appendix

\section[Wavefunctions of the $2D$-dimensional harmonic oscillator]{Wavefunctions of the $\boldsymbol{2D}$-dimensional harmonic oscillator}\label{appendixA}

The purpose of this appendix is to derive the explicit form of the harmonic oscillator wavefunctions (\ref{eq:wf-osc}).

On inserting (\ref{eq:wf-osc}) in the Schr\"odinger equation (\ref{eq:SE-osc}), the latter separates into $D-1$ angular equations
\begin{gather}
    \left\{- d_{\theta_{\nu}}^2 - [(2D-2\nu-1) \cot \theta_{\nu} - \tan \theta_{\nu}] d_{\theta_{\nu}} +
       \frac{C_{\nu+1}}{\sin^2 \theta_{\nu}} + \frac{p_{\nu}^2}{\cos^2 \theta_{\nu}} - C_{\nu} \right\}
       \Theta_{\nu}(\theta_{\nu}) = 0, \nonumber\\
  \nu=1, 2, \ldots, D-1,
   \label{eq:angular}
\end{gather}
and a radial equation
\begin{gather}
  \left(- d_R^2 - \frac{2D-1}{R} d_R + \frac{C_1}{R^2} + R^2 - E^{\rm osc}\right) {\cal L}(z) = 0.
  \label{eq:radial}
\end{gather}
Here $C_1,C_2, \ldots,C_{D-1}$ are $D-1$ separation constants, while $C_D$ is def\/ined by
\begin{gather}
  C_D = p_D^2. \label{eq:C-D}
\end{gather}
In the following, we are going to show that there does exist a solution to the whole set of $D$ equations (\ref{eq:angular}) and (\ref{eq:radial}) such that all the separation constants $C_{\nu}$, $\nu=1, 2, \ldots, D-1$, are nonnegative.

Let us start by solving the angular equation (\ref{eq:angular}) corresponding to the variable $\theta_{\nu}$ in terms of $p_{\nu}$ and $C_{\nu+1}$. The ansatz
\begin{gather*}
  \Theta_{\nu}(\theta_{\nu}) = (\cos \theta_{\nu})^{a_{\nu} - \frac{1}{2}} (\sin \theta_{\nu})^{b_{\nu} -
  \frac{1}{2}} F_{\nu}(u_{\nu}), \qquad u_{\nu} = \cos^2 \theta_{\nu},
\end{gather*}
transforms it into the hypergeometric dif\/ferential equation \cite{gradshteyn}
\begin{gather}
  \left\{u_{\nu} (1 - u_{\nu}) d_{u_{\nu}}^2 + [\gamma - (\alpha+\beta+1) u_{\nu}] d_{u_{\nu}} - \alpha \beta
  \right\} F_{\nu}(u_{\nu}) = 0  \label{eq:hyper}
\end{gather}
provided we choose the constants $a_{\nu}$ and $b_{\nu}$ in such a way that
\begin{gather}
  \left(a_{\nu} - \tfrac{1}{2}\right)^2 = p_{\nu}^2, \qquad \left(b_{\nu} - \tfrac{1}{2}\right) \left(b_{\nu} + 2D
  - 2\nu - \tfrac{5}{2}\right) = C_{\nu+1}.  \label{eq:quadratic}
\end{gather}
In (\ref{eq:hyper}), $\alpha$, $\beta$, and $\gamma$ are given by
\begin{gather*}
  \alpha = \tfrac{1}{2} (a_{\nu} + b_{\nu} + D - \nu - 1 + \Delta_{\nu}),\! \qquad \beta = \tfrac{1}{2} (a_{\nu} +
  b_{\nu} + D - \nu - 1 - \Delta_{\nu}),\! \qquad \gamma = a_{\nu} + \tfrac{1}{2},
\end{gather*}
where
\begin{gather}
  \Delta_{\nu} = \sqrt{(D-\nu)^2 + C_{\nu}}.  \label{eq:delta}
\end{gather}
There are altogether four solutions to the two quadratic equations (\ref{eq:quadratic}), which may be written as
\begin{gather}
  a_{\nu} = \tfrac{1}{2} + \epsilon |p_{\nu}|, \qquad b_{\nu} = - \left(D - \nu - \tfrac{3}{2}\right) + \epsilon'
  \Delta_{\nu+1}, \qquad \epsilon, \epsilon' = \pm 1.  \label{eq:a-b-bis}
\end{gather}
Consequently, we get
\begin{gather}
  \alpha = \tfrac{1}{2}(1 + \epsilon |p_{\nu}| + \epsilon' \Delta_{\nu+1} + \Delta_{\nu}), \qquad \beta =
  \tfrac{1}{2}(1 + \epsilon |p_{\nu}| + \epsilon' \Delta_{\nu+1} - \Delta_{\nu}), \nonumber\\
   \gamma = 1 + \epsilon
  |p_{\nu}|.  \label{eq:alpha}
\end{gather}

The general solution of the dif\/ferential equation (\ref{eq:hyper}) may be written down as
\begin{gather*}
  F_{\nu}(u_{\nu}) = A \, {}_2F_1(\alpha, \beta; \gamma; u_{\nu}) + B u_{\nu}^{1-\gamma}
  \, {}_2F_1(\alpha-\gamma+1, \beta-\gamma+1; 2-\gamma; u_{\nu}),
\end{gather*}
where $A$ and $B$ are two constants to be determined so that the angular function $\Theta_{\nu}(\theta_{\nu})$ be physically acceptable, i.e., vanish for $\theta_{\nu} \to 0$ and $\theta_{\nu} \to \frac{\pi}{2}$. On considering the four possibilities for the pair $(\epsilon, \epsilon')$ in (\ref{eq:a-b-bis}) and (\ref{eq:alpha}) successively, we arrive at a single solution corresponding either to $\epsilon = +1$, $\epsilon' = +1$, $B = 0$, $\beta = - n_{\nu}$ ($n_{\nu} \in \N$) or to $\epsilon = -1$, $\epsilon' = +1$, $A = 0$, $\beta-\gamma+1 = - n_{\nu}$ ($n_{\nu} \in \N$). It can be expressed in terms of a Jacobi polynomial \cite{gradshteyn} as in equation (\ref{eq:theta}), where
\begin{gather}
  a_{\nu} = |p_{\nu}| + \tfrac{1}{2}, \qquad b_{\nu} = - (D-\nu-\tfrac{3}{2}) + \Delta_{\nu+1}, \qquad
  n_{\nu}=0, 1, 2, \ldots,  \label{eq:a-b-ter}
\end{gather}
while the separation constant $C_{\nu}$ must satisfy the equation
\begin{gather}
  \Delta_{\nu} = 2n_{\nu} + |p_{\nu}| + \Delta_{\nu+1} + 1  \label{eq:delta-rec}
\end{gather}
with $\Delta_{\nu}$ def\/ined in (\ref{eq:delta}).

To obtain a solution to the whole set of $D-1$ angular equations (\ref{eq:angular}), as expressed in equation (\ref{eq:theta-bis}), it only remains to solve the recursion relation (\ref{eq:delta-rec}) for $\Delta_{\nu}$ with the starting value $\Delta_D = |p_D|$ corresponding to (\ref{eq:C-D}). The results for $\Delta_{\nu}$ and $C_{\nu}$, $\nu=1, 2, \ldots,D-1$, read
\begin{gather*}
  \Delta_{\nu} = 2n_{\nu} + 2n_{\nu+1} + \cdots +2n_{D-1} + |p_{\nu}| + |p_{\nu+1}| + \cdots + |p_D| + D - \nu
\end{gather*}
and
\begin{gather}
  C_{\nu}  = (2n_{\nu} + 2n_{\nu+1} + \cdots +2n_{D-1} + |p_{\nu}| + |p_{\nu+1}| + \cdots + |p_D|) \nonumber\\
\phantom{C_{\nu}  =}{}  \times (2n_{\nu} + 2n_{\nu+1} + \cdots +2n_{D-1} + |p_{\nu}| + |p_{\nu+1}| + \cdots + |p_D| + 2D -
        2\nu),
  \label{eq:C-nu}
\end{gather}
respectively. As a consequence, $b_{\nu}$ in (\ref{eq:a-b-ter}) can be rewritten as in equation (\ref{eq:a-b}). This completes the proof of equations (\ref{eq:theta-bis}) to (\ref{eq:a-b}).

Turning now ourselves to the radial equation (\ref{eq:radial}), we note from (\ref{eq:C-nu}) that $C_1$ can be written as
\begin{gather*}
  C_1 = 4j (j+D-1)
\end{gather*}
in terms of $j$ def\/ined in (\ref{eq:j}). Finally, it is straightforward to show that the physically acceptable solutions vanishing for $r$ (or $z$) going to zero and inf\/inity are given by (\ref{eq:L}) and correspond to the eigenvalues (\ref{eq:E-osc}).

\pdfbookmark[1]{References}{ref}
\LastPageEnding

\end{document}